\documentstyle[11pt,newpasp,twoside,epsf]{article}
\newcommand{\xt}[1]{\mbox{$\times 10^{#1}$}}
\newcommand{\x}[1]{\hspace*{#1mm}}

\begin{document}

\title{Grains in Photo-Ionized Environments}

\author{P.A.M. van Hoof$^{1,2}$, J.C. Weingartner$^1$, P.G. Martin$^{1}$, K. Volk$^{3}$,
and G.J.~Ferland$^{2}$}
\affil{
$^{1}$CITA, 60 St. George Street, Toronto, ON M5S~3H8, Canada\\
$^{2}$University of Kentucky, 177 CP Building, Lexington, KY 40506, USA\\
$^{3}$University of Calgary, 2500 University Dr. NW, Calgary, AB T2N~1N4, Canada
}

\begin{abstract}
Ever since the pioneering study of Spitzer (1948), it has been widely
recognized that grains play an important role in the heating and cooling of
photo-ionized environments. This includes the diffuse ISM, as well as H II
regions, planetary nebulae, and photo-dissociation regions.
A detailed code is necessary to model grains in a photo-ionized medium since
the interactions of grains with their environment include a host of
microphysical processes, and their importance can only be judged by performing
a complete simulation. In this paper we will use the spectral synthesis code
Cloudy for this purpose. A comprehensive upgrade of the grain model has been
recently incorporated in Cloudy, and certain aspects of this upgrade will be
discussed. Special emphasis will be on the new grain charge model. We will
consider in detail the physics of grains in both ionized and neutral
environments, and will present a calculation of photo-electric heating and
collisional cooling rates for a range of physical conditions and grain
materials and for a range of grain sizes (including a realistic size
distribution). We conclude with a brief discussion of the problems currently
hampering progress in this field. The new grain model will be used to model
the silicate emission in the Ney-Allen nebula, and will help us better
understand the nature of the grains in that part of the Orion complex.
\end{abstract}

\section{Introduction}

Grains are ubiquitous in the interstellar medium (ISM), and they can be
detected either directly through their far-infrared emission or indirectly
through extinction or polarization studies. Despite the vast number of
observations, many questions regarding grain composition and grain physics
remain unanswered. Further study is therefore required, and detailed models
are needed to interpret the results. Ever since the pioneering study of
Spitzer (1948), it has been widely recognized that grains play an important
role in the heating and cooling of the diffuse ISM (see also the more recent
studies by Bakes \& Tielens 1994, and Weingartner \& Draine 2001a, hereafter
WD). Grains also play an important role in the physics of H\,{\sc ii} regions
and planetary nebulae (e.g., Maciel \& Pottasch 1982, Baldwin et al.\ 1991,
hereafter BFM, Borkowski \& Harrington 1991, and Volk, these proceedings) and
photo-dissociation regions (PDR's, e.g., Tielens \& Hollenbach 1985).

The interactions of grains with their environment include a host of
microphysical processes, and their importance and effects can only be judged
by including all of these processes. This can, in turn, only be done with a
complete simulation of the environment. In this paper we consider in detail
the physics of grains in both ionized and neutral environments, and model
these with the spectral synthesis code Cloudy. We will present a calculation
of photo-electric heating and collisional cooling rates for a range of
physical conditions and grain materials and for a range of grain sizes
(including a realistic size distribution). A comparison of our results with
benchmark calculations using the WD code will also be presented.

\section{The Photo-Ionization Code Cloudy}

Cloudy is a well known and widely used photo-ionization code which is publicly
available at {\tt http://www.pa.uky.edu/$^\sim$gary/cloudy}. This code is not
only useful for modeling fully ionized regions, but calculations can also be
continued into the PDR. In order to make such a calculation realistic, the
presence of a detailed grain model is required. The first grain model was
introduced to Cloudy in 1990 to facilitate more accurate modeling of the Orion
nebula (for a detailed description see BFM). In subsequent years, this model
has undergone some revisions and extensions, but remained largely the same.

Recently, Cloudy has undergone several major upgrades, described in Ferland
(2000a), Ferland (2000b), and van Hoof, Martin, \& Ferland (2000). Some of these
recent upgrades were aimed at improving the accuracy and versatility of Cloudy
as a PDR code. These comprise an upgrade in the collision strengths of the
[C\,{\sc i}] and [O\,{\sc i}] infrared fine-structure lines, the inclusion of
a full CO model, and a comprehensive upgrade of the grain model. The latter
was necessary for two reasons. First, the discovery of crystalline silicates
in stellar outflows (e.g., Waters et al.\ 1996), and other detailed
observations of grain emission features by the Infrared Space Observatory
(ISO), meant that the code had to become much more flexible to allow such
materials to be included in the modeling. Second, even before the ISO mission
it had already become clear that the photo-electric heating and collisional
cooling of the gas surrounding the grains is dominated by very small grains
(possibly consisting of polycyclic aromatic hydrocarbons or PAH's). The
charging of very small grains could not be modeled very accurately with the
original grain model (grain physics becomes increasingly non-linear as a
function of charge for smaller grains and the average potential approach of
the original grain model eventually breaks down for molecule-sized grains). In
view of these facts we have undertaken a comprehensive upgrade of the grain
model in Cloudy. The two main aims were to make the code more flexible and
versatile, and to make the modeling results more realistic. The new grain
model is included in version 96 of Cloudy.\footnote{A beta release is
currently available at {\tt http://nimbus.pa.uky.edu/cloudy/cloudy\_\,96.htm}}

\section{Resolving the Grain Size Distribution}
\label{resolv}

\begin{figure}
\centerline{\epsfxsize=0.45\columnwidth\epsfbox[62 305 479 653]{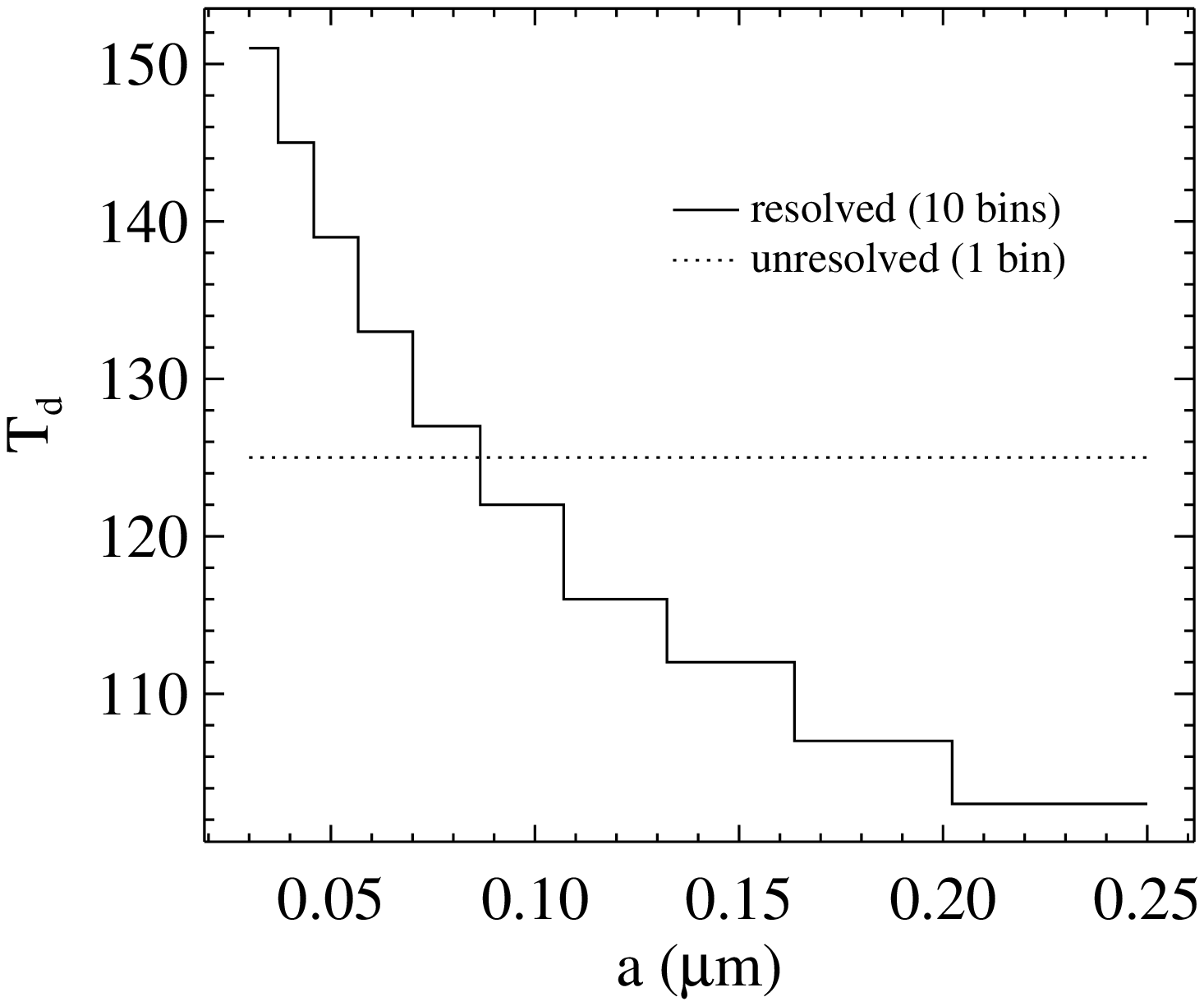}
\hspace{2em}\epsfxsize=0.45\columnwidth\epsfbox[62 305 479 653]{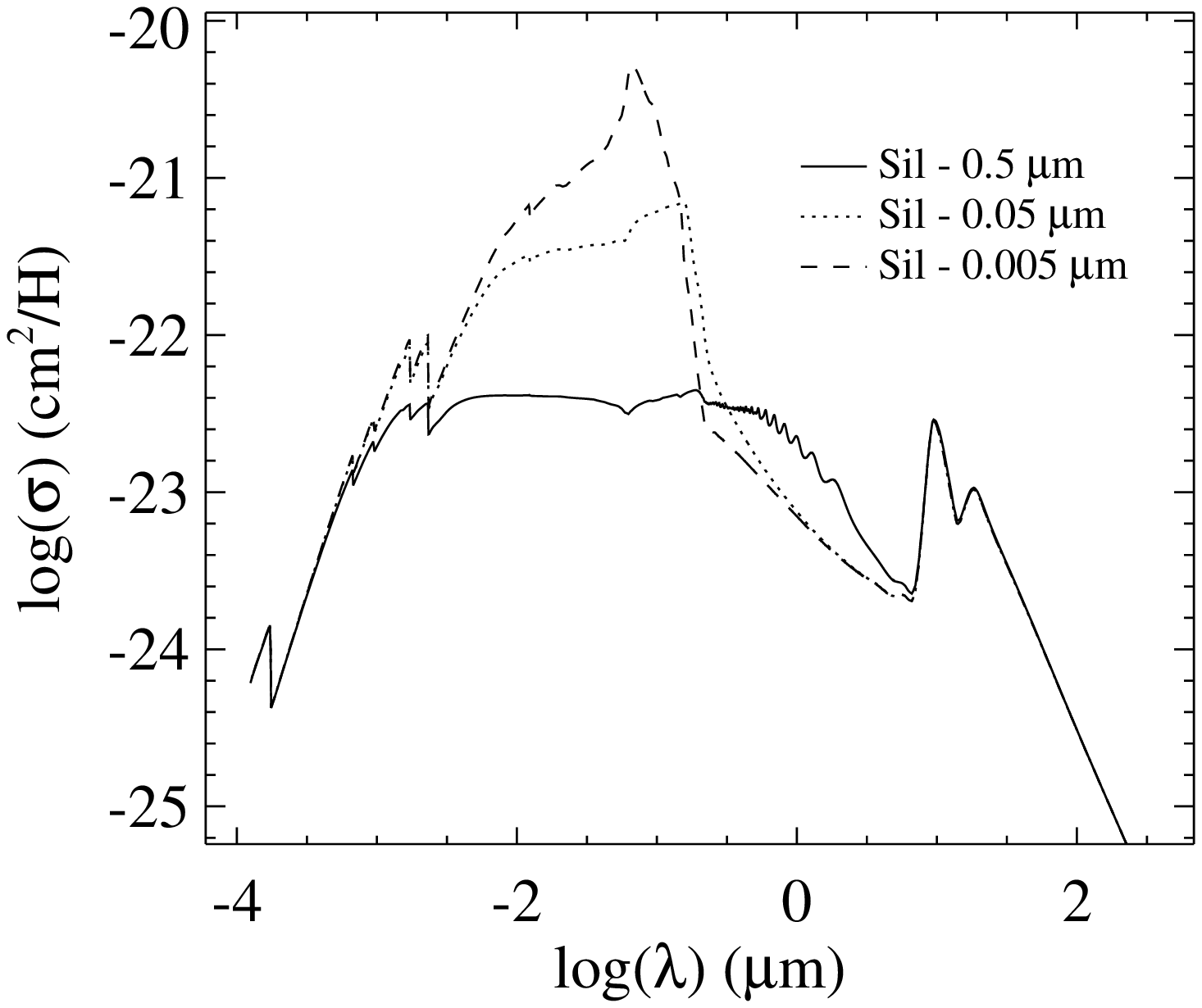}}
\caption{{\bf Left Panel:} The average grain temperature in each size bin
plotted as a function of grain size for astronomical silicate in the innermost
zone of our Orion model (BFM). The solid line shows the results for a resolved
size distribution, and the dotted line the results for an unresolved
distribution. {\bf Right Panel:} The absorption cross section for astronomical
silicate (Draine \& Lee 1984, Martin \& Rouleau 1991) for three single sized
grains. The dust-to-gas ratio is the same for all three species and the cross
sections are normalized per hydrogen nucleus in the plasma.
\label{plot:sd}}
\end{figure}

In the original grain model, opacities for a handful of grain species were
hard-wired in the code. Furthermore, only a single size bin would be used for
the entire size distribution. This is not a very good approximation since most
grain properties depend strongly on size, e.g., the grain opacities (see the
right panel of Figure~\ref{plot:sd}). This approach was nevertheless adhered
to in BFM because of computational restrictions.

Resolving the size distribution into many small bins improves the modeling in
several ways. First, an equilibrium temperature can be calculated for each bin
separately (see the left panel of Figure~\ref{plot:sd}). This is important
since grain emissions are a strongly non-linear function of temperature.

More importantly, resolving the size distribution also enables other
improvements of the grain treatment: stochastic heating can be treated
correctly for the smallest grains in the size distribution (see
\S~\ref{qheating}), and the calculations yield much more accurate results for
the total photo-electric heating and collisional cooling rates of the gas by
the grains (see \S~\ref{physics}).

To improve the model, we have implemented the following
changes:

1 -- We have included a Mie code for spherical particles in Cloudy. Assuming
that the grains are homogeneous spheres with a given complex refractive index
(optical constant) one can use Mie theory (Mie 1908) to calculate the
absorption and scattering opacity. This has to be done separately for every
wavelength since the refractive index depends on wavelength. Good overviews of
Mie theory can be found in van de Hulst (1957), and Bohren \& Huffman (1983).
Our Mie code is based on the program outlined in Hansen \& Travis (1974) and
references therein. The optical constants needed to run the code are read from
a separate file. This allows greater freedom in the choice of grain species.
Files with optical constants for a range of materials are included in the
Cloudy distribution. However, the user can also supply optical constants for a
completely different grain type.


2 -- It is possible to use arbitrary grain size distributions. The user can
either choose one of a range of preset functions (with numerous free
parameters), or supply the size distribution in the form of a table.
Single-sized grains can also be treated.

3 -- The size distribution can be resolved in an arbitrary number of size bins
(set by the user), and the absorption and scattering opacities and all the
physical parameters (charge, temperature, etc.) are calculated for each bin
separately.

\subsection{The Dust Emission Spectrum in Orion}

\begin{figure}
\centerline{\epsfxsize=0.45\columnwidth\epsfbox[84 305 479 653]{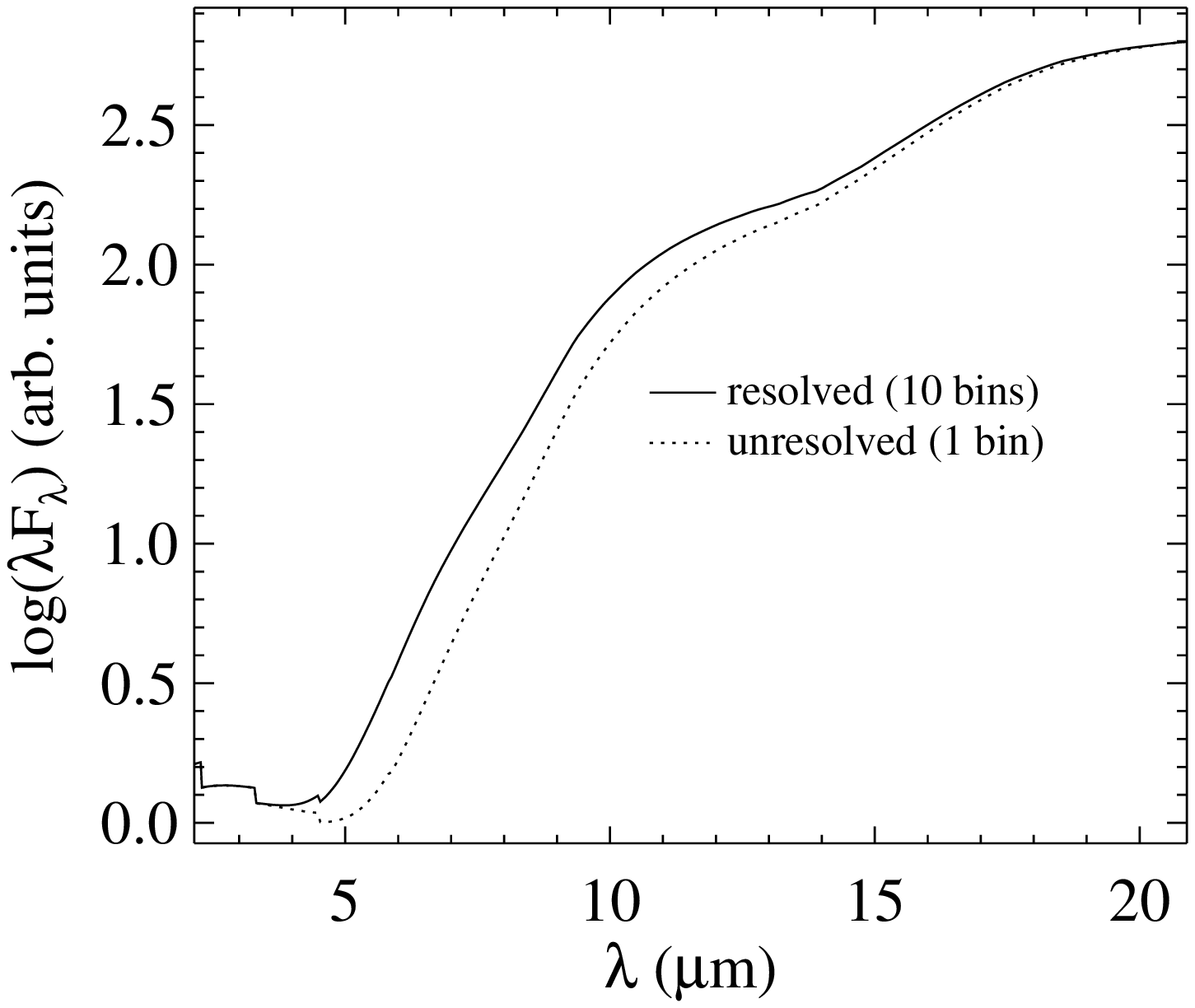}
\hspace{2em}\epsfxsize=0.45\columnwidth\epsfbox[84 305 479 653]{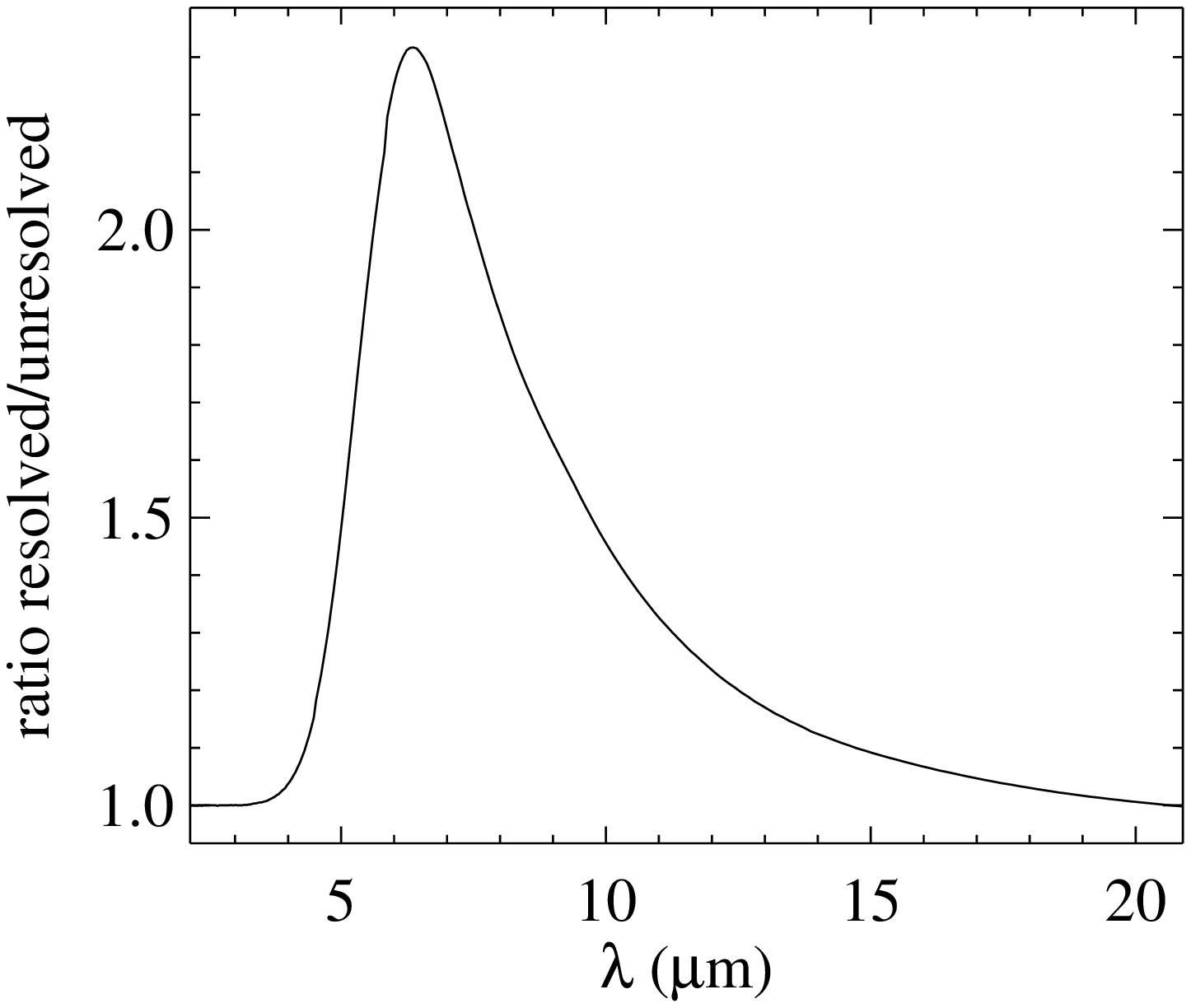}}
\caption{{\bf Left Panel:} Model predictions for the grain emission from the
face of the Orion molecular cloud complex illuminated by $\theta^1$ Ori~C,
using resolved (solid line) and unresolved (dotted line) grain size
distributions. Stochastic heating was turned off in these models to highlight
the effect of resolving the size distribution. The flux in the center of the
10-$\mu$m feature has gone up by 45\%, in the blue wing by more than a factor
of two! {\bf Right Panel:} The flux-ratio of the resolved and unresolved
calculations.\label{effect:sd}}
\end{figure}

In Figure~\ref{effect:sd} we display the dust emission from the face of the
Orion blister illuminated by $\theta^1$ Ori~C, using the BFM model with
resolved and unresolved grain size distributions. These calculations include
grains in the ionized region, the PDR, and fully molecular regions. Stochastic
heating effects were turned off for these calculations to highlight the effect
of resolving the size distribution. These effects would have been small anyway
since this model is based on the BFM Orion size distribution which does not
contain grains smaller than 30~nm. One can see that resolving the size
distribution has a noticeable effect on the emitted spectrum since the grain
emissivity depends strongly on size.

\section{Stochastic Heating of Small Grains}
\label{qheating}

It is well known that in conditions where the cooling time of the grain is
shorter than or comparable to the average time between two significant heating
events, a stochastic treatment of the grain temperature is necessary. This
effect is important for grains smaller than roughly 20~nm (most notably for
PAH's), and/or in regions where the photon density is very low (e.g., the
diffuse ISM). The importance of this effect was first suggested by Greenberg
(1968), and the topic has evolved considerably since (e.g., Desert el al.\
1986, Dwek 1986), mainly focusing on efficient numerical techniques for
calculating the temperature distribution of the grains. A brief history can be
found in Guhathakurta \& Draine (1989, hereafter GD).

Code implementing stochastic treatment of PAH's was already included in a
revision to the old grain model. However, that code only worked with the two
PAH species that were hardwired in Cloudy, and therefore was only infrequently
used. In order to obtain accurate modeling results, the code should work on
very small grains in any size distribution (which is only possible now that
the size distribution has been resolved), and should include the effects
automatically when they have a noticeable effect on the emitted
spectrum\footnote{In Cloudy version 96beta2 stochastic heating effects still
need to be turned on explicitly by the user. Once the numerical stability of
the new algorithm has been sufficiently validated, the situation will be
reversed and stochastic heating effects will be included automatically unless
explicitly turned off by the user.}. To achieve these goals, the stochastic
heating code has been extensively rewritten for the new grain model. It now
works efficiently with all grain types and sizes, and under all conditions.

The original code in Cloudy was based on the algorithm of GD. The aim of
this algorithm is to calculate the probability distribution of grain
temperatures (or equivalently: grain enthalpies). When the code was upgraded
for the current release of Cloudy, certain improvements were added to this
algorithm for better performance.

\begin{figure}[th]
\centerline{\epsfxsize=0.5\columnwidth\epsfbox[33 304 510 653]{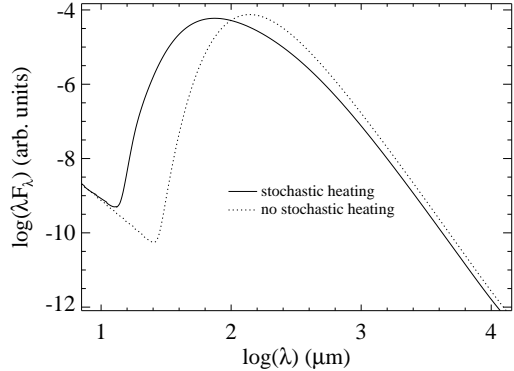}}
\caption{Model predictions for the emission from a 5~nm silicate grain in
typical diffuse ISM conditions. The dotted line shows the model assuming that
all grains have the same average temperature (i.e., stochastic heating effects
are turned off). The solid line shows the model including stochastic effects.
The model is for a hydrogen column density of $10^{20}$~cm$^{-2}$; the
continuum at 10~$\mu$m is the interstellar radiation field from nearby stars.
\label{effect:qh}}
\end{figure}

The first difference with the work of GD is that we assumed $\lambda_{\rm
cutoff} = \infty$ and dropped the term containing the integrated incident flux
for wavelengths longer than $\lambda_{\rm cutoff}$ from the energy balance.
The quantity $\lambda_{\rm cutoff}$ has no physical meaning, and was merely a
numerical invention introduced by GD to avoid dealing with zero temperature
grains. Our algorithm incorporates a different and more natural solution for
that problem. This improved convergence of the probability distribution
dramatically and helped us to avoid calculating grain enthalpy bins with
extremely low probabilities. The algorithm of GD was rewritten in such a way
that it is no longer necessary to set up the grain enthalpy grid {\it a
priori}, but use an adaptive stepsize algorithm instead. This makes the
algorithm much more flexible and efficient, while still guaranteeing proper
convergence. The code can automatically detect when stochastic heating might
be important, or when it is safe to simply use the equilibrium temperature for
calculating the emitted spectrum. To highlight the importance of stochastic
heating, we show in Figure~\ref{effect:qh} the spectrum emitted by a 5~nm
silicate grain in typical diffuse ISM conditions with stochastic heating
effects either included or not.

\section{Changes to the Grain Physics}
\label{physics}

We have modified certain aspects of the grain physics following the discussion
in WD. Below we highlight certain aspects of these changes. A detailed
description will be presented in a forthcoming paper (van Hoof et al.,
in preparation).

1 -- We include the bandgap between the valence and conduction bands in our
potential well model for silicates. This change only affects the results for
negatively charged grains ($Z_{\rm g} \leq -1$).

2 -- Reduction of the potential barrier for negatively charged grains is
included using an analytic fit to numerical calculations. Two effects are
important here: quantum tunneling and the Schottky effect. Quantum theory
predicts that an electron with insufficient energy to overcome a barrier still
has a finite chance of tunneling through. This effect has been modeled using
the WKB approximation which gives a simple analytic expression for the
tunneling probability for a barrier of given width and height. Quantum
tunneling is only important for small grains. For large grains the Schottky
effect will dominate, which describes the lowering of the potential barrier by
an image potential in the grain. This effect has been accurately modeled by
Draine \& Sutin (1987).

We will approximate both effects by assuming that the barrier is effectively
reduced in height from $ - ( Z_{\rm g} + 1 ) e^2/(4\pi\epsilon_0 a)$ to
$-E_{\rm min}$. The magnitude of the combined tunneling/Schottky effect was
calculated by WD. However, the fitting function they used has the wrong
limiting behavior for large grains where it should asymptotically approach the
classical Schottky expression. We therefore repeated these calculations using
the same assumptions, but adopted a different fitting function that does
exhibit the correct limiting behavior:
\begin{equation}
\label{thres:new}
    E_{\rm min} = -  \, \theta_\nu \frac{e^2}{4\pi\epsilon_0 a}
    \left[ 1  - \frac{0.3}{ ( a / {\rm nm} )^{0.45} \, \nu^{0.26}} \right],
\end{equation}
where $\theta_\nu[\nu = - ( Z_{\rm g} + 1 ) \, ]$ describes the Schottky
effect and is defined in Draine \& Sutin (1987). The term in square brackets
describes the quantum-mechanical correction. This change only affects the
results for grains with $Z_{\rm g} < -1$.

3 -- The treatment of the photo-electric effect has been improved, following
the discussion in WD. This includes new expressions for the photo-electric
yield and the energy distribution of ejected electrons.

4 -- The treatment of electron sticking probabilities has been updated, again
following WD. Especially for very small grains the sticking efficiency has
been substantially lowered to obtain better agreement with laboratory studies of
molecules. This has an important impact on the photo-electric heating rate of
the gas since the electron recombination rate has to be matched by electron
loss processes to preserve the charge balance. The loss processes are usually
dominated by the photo-electric effect.

5 -- Certain physical constants have been updated. Most notably the work
function for graphite has been lowered. This results in an increased
photo-electric heating rate of the gas since less of a potential barrier needs
to be overcome to ionize graphite grains.

6 -- The treatment of collisional heating or cooling of the grains and the gas
has been improved.

Our treatment deviates from the WD code in two ways. Most importantly, we use
a different grain charge model, which will be discussed in more detail in
\S~\ref{hybrid} Secondly, we use slightly different physics for collisions
between ions and grains. This only gives rise to very small differences at the
1 -- 2\% level when compared to the WD treatment.

\subsection{The New Hybrid Grain Charge Model}
\label{hybrid}

The original grain model in Cloudy (which we will call the average grain
potential model) is described in BFM. In that model an average grain potential
is calculated by finding the potential for which the charge gain rate exactly
matches the loss rate. This method was first proposed by Spitzer (1948), and
is an excellent approximation for large grains. However, it is now clear that
photo-electric heating and collisional cooling of the gas are dominated by
very small grains. For such grains the average grain potential approximation
does not work very well because grain physics becomes increasingly non-linear
as a function of charge for smaller grain sizes. This fact, combined with the
fact that grain charges are quantized, has led to a new approach where the
charge distribution is fully resolved, and heating and cooling rates are
calculated for each charge state separately (see e.g., WD). This ensures
accurate results, but leads to an appreciable increase in computational
overhead. This is especially the case for large grains since the width of the
charge distribution increases with grain size. Hence the paradoxical situation
arises that most of the computing time is spent on grains which contribute
least to heating and cooling, and which are also the grains for which the
average grain potential model works best!

In this paper we present a hybrid grain potential model which saves most of
the computational speed of the original average grain potential model, but
nevertheless gives sufficient accuracy when compared to fully resolved charge
distribution calculations. The basic philosophy is that for very small grains
($a < 1$~nm) only a few charge states have a significant population. Hence we
adopt the $n$-charge state approximation, in which all grains are treated by
using exactly $n$ contiguous charge states, independent of size. The higher
$n$ is, the more accurate the results will be (exactly how accurate will be
discussed in \S~\ref{validating}). The default for Cloudy calculations is $n =
2$, but the user can request a larger number if higher precision is desired.

Since the $n$-charge state model does not fully resolve the charge
distribution, a different algorithm from WD is needed to calculate the
fractional populations $f_i$ of each of the $n$ charge states $Z_i \equiv Z_1
+ i -1$. These populations must first of all obey the following normalization:
\begin{equation}
\label{c1}
   \sum_{i=1}^n f_i = 1.
\end{equation}
Secondly, we require that the electron gain rates $J_i^-$ and electron loss
rates $J_i^+$ summed over all charge levels match exactly:
\begin{equation}
\label{c2}
   \sum_{i=1}^n f_i ( J_i^+ - J_i^- ) = 0,
\end{equation}
similar to the average grain potential model. Eqs.~\ref{c1} and \ref{c2} are
sufficient to determine the charge state populations if $n = 2$, but for $n >
2$ we need additional equations. These equations need to satisfy the following
constraints. First, the resulting level populations should always be greater
or equal zero. Second, the level populations should change continuously when
the electron gain and loss rates change continuously. Third, the level
populations should asymptotically approach the results from fully resolved
calculations for increasing values of $n$. We have adopted the following
algorithm:

1 -- The $n$ charge states are split up in two groups of $n-1$ contiguous
charge states. The first group contains charge states $[Z_1,Z_{n-1}]$, and the
second $[Z_2,Z_n]$. The value for $Z_1$ is determined iteratively (see step~4).

2 -- The relative level populations in the first group $f_i^1$ are determined
using an algorithm very similar to the one used in fully resolved
calculations, i.e., assume $f_1^1 =1$, calculate $f_2^1 = f_1^1 J_1^+ / J_2^-$,
$f_3^1 = f_2^1 J_2^+ / J_3^-$, etc.\footnote{Note that this procedure is not
correct for charge transfer with multiply charged ions. In order to avoid
having to solve a full set of linear equations, we will approximate this
process as multiple single-charge-transfer events. The resulting errors are
expected to be small as collision rates for multiply charged ions are usually
quite low because their velocities are small compared to electrons. The
collision rates are normally suppressed even further by the positive grain
charge.}, and then re-normalize to $\sum_{i=1}^{n-1} f_i^1 = 1$. We use an
analogous procedure for the populations $f_i^2$ of the second group for $i \in
[2,n]$.

3 -- Determine for both groups the net charging rate
\begin{equation}
\label{c3a}
   J_k = \sum_{i=k}^{n-2+k} f_i^k ( J_i^+ - J_i^- ) \x{3} (k = 1,2).
\end{equation}

4 -- Iterate $Z_1$ and repeat steps 1 -- 3 until $J_1 \times J_2 \leq 0$. Then
find $0 \leq \alpha \leq 1$ such that
\begin{equation}
\label{c3b}
   \alpha J_1 + (1 - \alpha) J_2 = 0.
\end{equation}

5 -- Determine the final charge state populations as follows
\begin{equation}
\label{c3c}
   f_i = \alpha \, f_i^1 + (1 - \alpha) \, f_i^2 \x{3} (\mbox{with } f_n^1 \equiv 0, f_1^2 \equiv 0).
\end{equation}
One can verify that this algorithm satisfies all constraints.

The hybrid grain potential model is efficient because it avoids the overhead
for large grains, while still giving accurate results for both small and large
grains. An added bonus is that most of the time an excellent initial estimate
for $Z_1$ can be derived from the previous zone, reducing the overhead even
further. The model works for very small grains because only few charge states
are populated and it can reconstruct the actual charge distribution. It works
for large grains because the grain potential distribution approaches a delta
function for increasing grain size (as opposed to the charge distribution
which becomes ever wider). Our method therefore asymptotically approaches the
average grain potential model, which we already know is very accurate for
large grains.

\subsection{Validating the Hybrid Grain Charge Model}
\label{validating}

In order to validate the new grain charge model, we calculated the
photo-electric heating and collisional cooling rates for a range of physical
conditions, two grain species, and a wide range of grain sizes (including a
realistic size distribution). We then compared these calculations with
benchmark results from the WD code, which fully resolves the charge
distribution.

We model the physical conditions with simple assumptions: the plasma only
contains hydrogen, the electron temperature and density are fixed at
prescribed values, the incident spectrum is assumed to be a blackbody (either
full in the warm ISM and ionized cases, or cut off at 13.6~eV in the cold ISM
and PDR cases). To allow for an accurate comparison, both codes used the same
optical constants for classical graphite and astronomical silicate (Draine \&
Lee 1984). In Cloudy the photo-electric heating by local diffuse emission was
switched off, as well as thermionic emissions and collisional cooling by
neutral hydrogen since these are not treated in the WD code. The WD code was
modified according to Eq.~\ref{thres:new}. Except for the grain charge model,
the treatment of the grain physics in Cloudy and the WD code are very similar.

We present results for selected single sized grains, as well as a realistic
reconstruction of the ISM grain size distribution (labeled A6) taken from
Weingartner \& Draine (2001b). In the ISM and H\,{\sc ii} region cases we used
a standard mix of graphite and silicate as prescribed by Weingartner \& Draine
(2001b)\footnote{Note that the A6 size distribution was derived by matching
the extinction curve using grain opacities defined in Li \& Draine (2001). In
this study we use the same size distribution, but use opacities from Draine \&
Lee (1984) instead, which is not consistent. This inconsistency is irrelevant
for our purposes.}. In the planetary nebula cases, we made separate models for
either graphite or silicate as these two species are not expected to coexist
spatially in the nebular material on theoretical grounds\footnote{Note that
spectroscopy of planetary nebulae by ISO has revealed a surprisingly large
number of cases that show both silicate and graphitic dust features. One
famous example is NGC~6302 (Molster et al.\ 2001). It is however usually
assumed that the two species reside in different parts of the nebula.}.

The physical parameters for the models are listed in Table~1; the results
of the calculation are presented in Tables~2, 3, and 4.

\begin{table}[t]
\begin{center}
\caption{\leftskip 13mm\setlength\hsize{121mm}
Physical parameters for the benchmark models. Symbols have their usual
meaning, $G$ is the intensity of the radiation field and $G_0 =
1.6\xt{-6}$~W\,m$^{-2}$, integrated between 6 and 13.6~eV, is the Habing
intensity. $T_{\rm c}$ is the color temperature of the radiation field.}
\begin{tabular}{lrrrrrr}
\hline
 & \multicolumn{2}{c}{ISM} & \multicolumn{2}{c}{H\,{\sc ii}} & \multicolumn{2}{c}{PN} \\
 & warm & cold & ionized & PDR & ionized & PDR \\
\hline
$T_{\rm c}$/kK             &  35 &   35 &   50 & 50 & 250 & 250 \\
log($G/G_0$)               &   0 &    0 &    5 &  5 &   5 &   5 \\
log($n_{\rm H}$/cm$^{-3}$) &   0 &    1 &    4 &  4 &   4 &   4 \\
log($n_{\rm e}$/cm$^{-3}$) &   0 & $-$2 &    4 &  1 &   4 &   1 \\
$T_{\rm e}$/kK             &   9 &  0.1 &    9 &  1 &  20 &   1 \\
\hline
\end{tabular}
\end{center}
\end{table}

\subsection{Discussion}

A summary of the comparison of the photo-electric heating and collisional
cooling rates from Tables~2, 3, and 4 is given in Table~\ref{summary}. In
general the results are in excellent agreement, with only a few outliers for
single sized grains in the $n=2$ and $n=3$ cases. The results for the size
distribution cases always agree to better than 25\%, even for $n=2$. This is
well within the accuracy with which we know grain physics to date. There are
still major uncertainties in the photo-electric yields and the sticking
efficiency for electrons, both of which have a strong effect on the
photo-electric heating rate. Also the work function and bandgap for
astrophysical grain materials are poorly known and can have a strong effect as
well. This is unfortunate since photo-electric heating and collisional cooling
in photo-ionized environments are important effects. All these uncertainties
mainly stem from the fact that the composition of interstellar grains is
still poorly known.

\begin{figure}[p]
\vspace*{-2mm}
\centerline{\epsfxsize=1.2\textwidth\epsfbox[55 62 532 749]{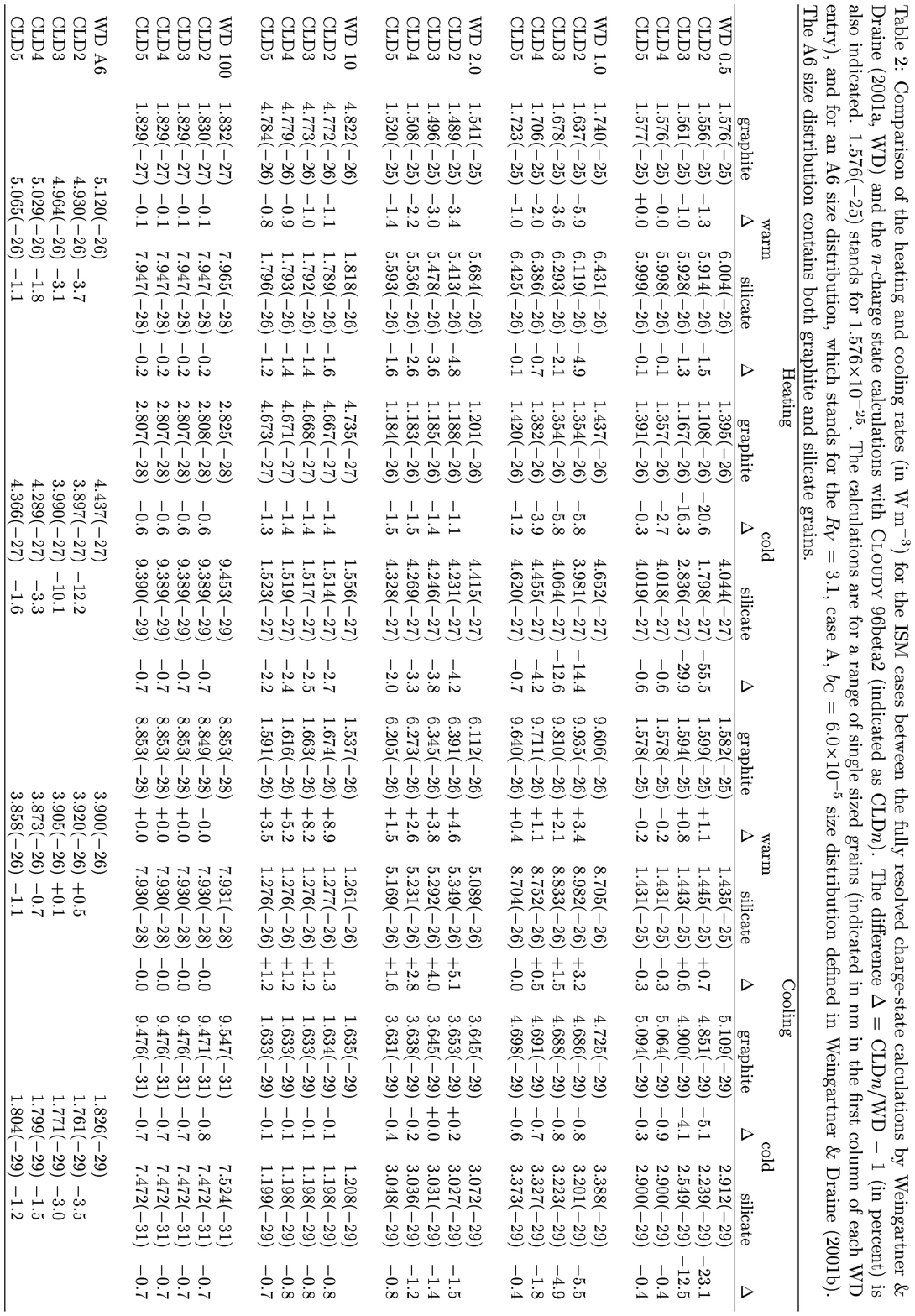}}
\end{figure}

\begin{figure}[p]
\vspace*{-2mm}
\centerline{\epsfxsize=1.2\textwidth\epsfbox[55 62 532 749]{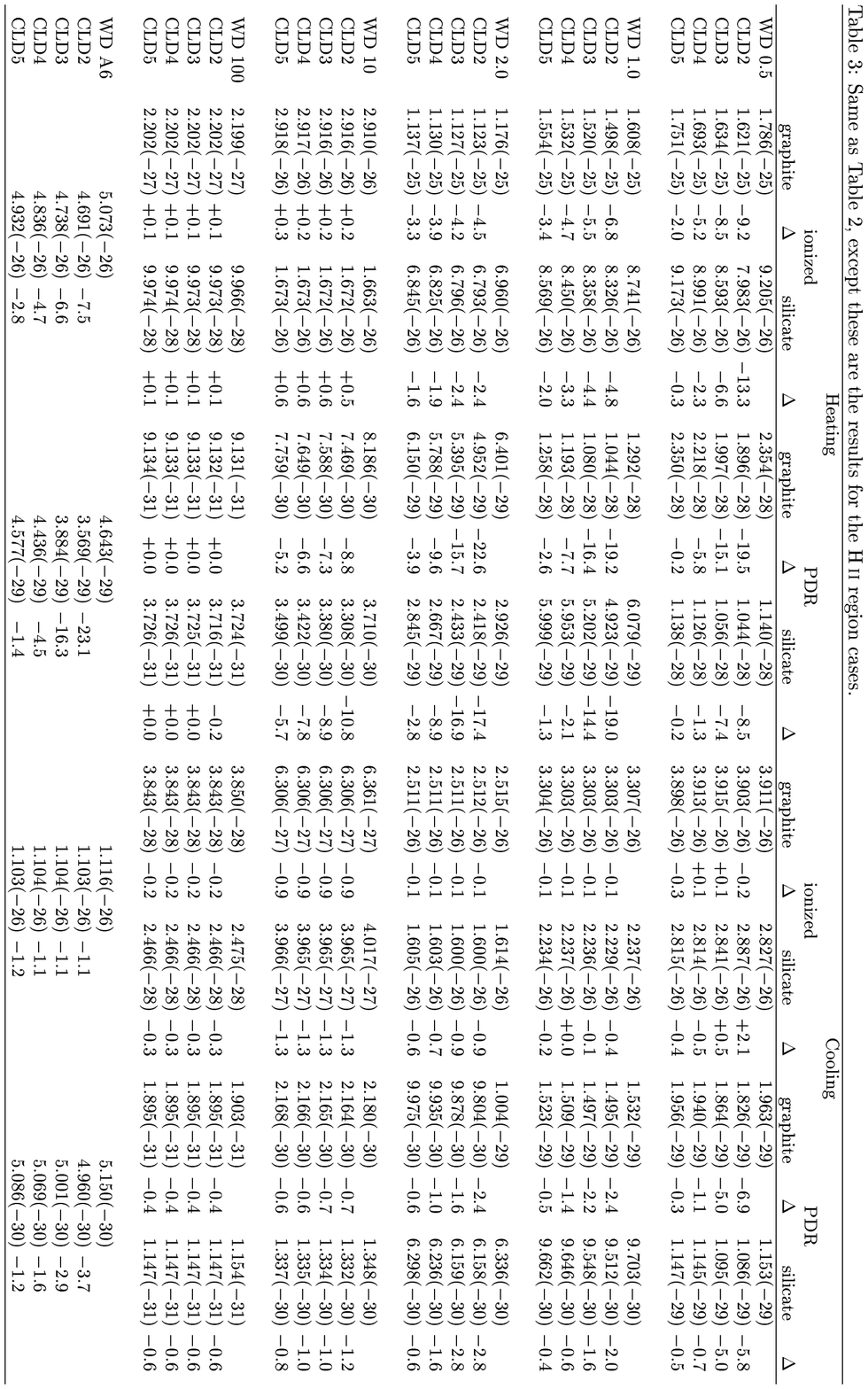}}
\end{figure}

\begin{figure}[p]
\vspace*{-2mm}
\centerline{\epsfxsize=1.2\textwidth\epsfbox[55 62 532 749]{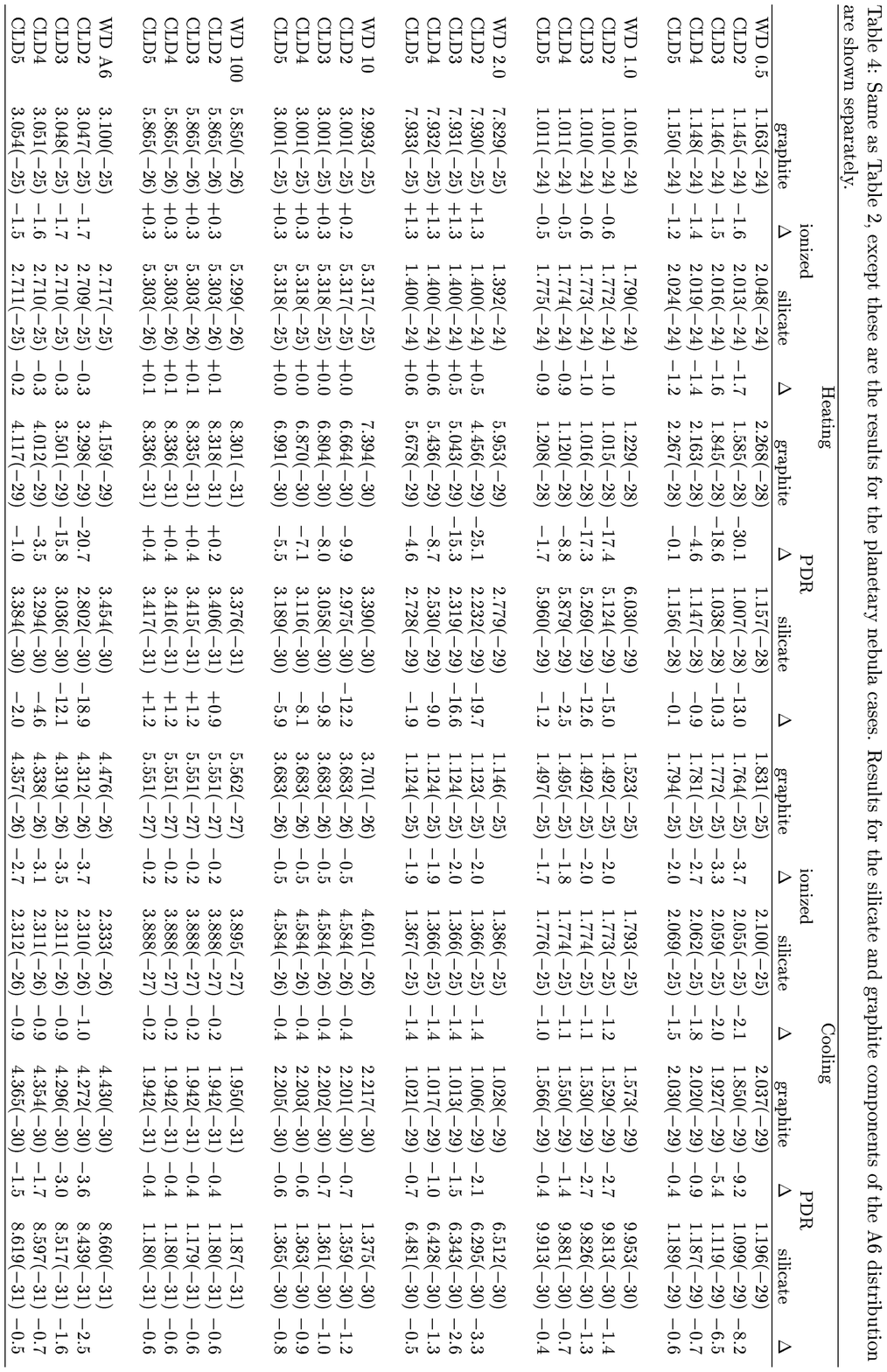}}
\end{figure}

From Table~\ref{summary} one can see that the collisional cooling rates
usually are in better agreement than the photo-electric heating rates
for a given set of physical parameters. It is furthermore clear that the
accuracy of the $n$-charge state approximation increases as $n$ increases,
as should be expected. Closer inspection of Tables~2, 3, and 4 for single
sized grains reveals that the largest errors in the $n=2$ and $n=3$ cases are
for the 0.5~nm grains, for $n=4$ for either 2~nm or 10~nm grains, and for $n=5$
for 10~nm grains, i.e., the grain size for which the errors are largest shifts
upwards for higher values of $n$. This is expected since the $n$-charge
state approximation will fully resolve the charge distribution of the smallest
grains for $n > 3$. Note that the results for the 100~nm grains are always
in excellent agreement, even when $n=2$, despite the fact that the actual
charge distribution is much wider than that.

The agreement between the Cloudy and the WD results is very satisfactory for
realistic size distributions, and should be sufficient for all realistic
astrophysical applications. Therefore the hybrid grain charge model presented
above (with $n=2$) will be the default for Cloudy modeling.

\setcounter{table}{4}

\begin{table}[t]
\begin{center}
\caption{\leftskip 0mm\setlength\hsize{\textwidth}Summary of the comparison
between the Cloudy $n$-charge state calculations (indicated by CLD$n$) and the
benchmark calculations with the WD code. All entries are differences CLD$n$/WD
$-$ 1 in percent.\label{summary}}
\begin{tabular}{lrrrrrrrr}
\hline
 & \multicolumn{4}{c}{Heating} & \multicolumn{4}{c}{Cooling} \\
 & \multicolumn{2}{c}{single size} & \multicolumn{2}{c}{size distr.} & \multicolumn{2}{c}{single size} & \multicolumn{2}{c}{size distr.} \\
 & median & worst & median & worst & median & worst & median & worst \\
\hline
CLD2 & $-3.03$ & $-55.5$ & $-9.85$ & $-23.1$ & $-0.77$ & $-23.1$ & $-3.05$ & $-3.7$ \\
CLD3 & $-2.44$ & $-29.9$ & $-8.34$ & $-16.3$ & $-0.72$ & $-12.5$ & $-2.27$ & $-3.5$ \\
CLD4 & $-1.40$ &  $-9.6$ & $-3.43$ &  $-4.7$ & $-0.65$ &  $+5.2$ & $-1.30$ & $-3.1$ \\
CLD5 & $-0.75$ &  $-5.9$ & $-1.45$ &  $-2.8$ & $-0.44$ &  $+3.5$ & $-1.21$ & $-2.7$ \\
\hline
\end{tabular}
\end{center}
\vspace*{-6pt}
\end{table}

\section{Outlook}

The rapid increase in computing power over the past decades has made it
possible now to build detailed grain models that can be run in a modest amount
of time. Computing power is therefore not a limiting factor anymore, and the
gaps in our knowledge of grain physics are almost exclusively what is
hampering progress at the moment. These gaps need to be filled in, both
through laboratory work, and observations of interstellar and circumstellar
grains. Some of the key areas have already been indicated above. They include
a better understanding of the exact chemical composition of the grains, and
the size distribution as a function of environment, as well as the
corresponding physical properties (electron and ion sticking efficiencies,
photo-electric yields, work functions, and bandgaps) for realistic
astronomical materials. Depletion of trace elements onto grains should also be
understood better since that has an important effect on the abundances of the
gas in which the grains have been formed and alters the emitted spectrum and
cooling rate of the gas. An improved model of the life cycle of dust
(including grain formation, coagulation, and destruction in the ISM, as well
as the formation of crystalline materials in circumstellar envelopes, which
was one of the great surprises from the ISO mission) should go hand in hand
with the above improvements. In the long run theory combined with laboratory
experiment should move towards building full quantum-mechanical models of very
small grains, which are now treated by heuristic, semi-classical methods.
Currently the complexity of such models is still computationally prohibitive.

We are currently using Cloudy to model the Ney-Allen nebula (Ney \& Allen
1969), which is situated close to $\theta^1$ Ori~D. The silicate emission from
this nebula was used (combined with data from long period variable stars) as a
template to define astronomical silicate (Draine \& Lee 1984). Now this effort
will come full circle by using the derived astronomical silicate data to model
the Ney-Allen nebula. This will be an important test of our understanding of
interstellar grains. The improvements presented above are important for this
effort since the silicate 10~$\mu$m feature is situated in the Wien tail of
the emission, and effects from resolving the size distribution (see
Fig.~\ref{effect:sd}) as well as temperature spiking of very small grains are
important. This work will be presented in a forthcoming paper (van Hoof et
al., in preparation).

\acknowledgements

We wish to acknowledge financial support by the National Science Foundation
through grant no.\ AST--0071180, and NASA through its LTSA program, NAG
5--3223. This research was also supported in part by the Natural Sciences and
Engineering Research Council of Canada. J.C.W. acknowledges support from an
NSF International Research Fellowship.

\end{document}